%% file: arxiv.tex
\documentclass{article} 
\usepackage{iclr2025_conference,times}

\iclrfinalcopy
\input{math_commands.tex}

\usepackage{hyperref}
\usepackage{url}
\usepackage{booktabs}
\usepackage{multirow}
\usepackage{array}
\usepackage{graphicx}
\usepackage{amsmath}
\usepackage{amsfonts}
\usepackage{amssymb}
\usepackage{caption}
\usepackage{adjustbox}
\usepackage{longtable}
\usepackage{adjustbox}
\usepackage{tcolorbox}
\usepackage{xcolor}   
\usepackage{booktabs}
\usepackage{array}
\usepackage{multirow}
\usepackage{siunitx}    
\usepackage[table]{xcolor}
\usepackage{caption}
\usepackage{fancyhdr}  

\newcolumntype{L}{>{\raggedright\arraybackslash}p{3.2cm}}
\newcolumntype{M}{>{\centering\arraybackslash}p{2.1cm}}
\newcolumntype{N}{>{\centering\arraybackslash}p{1.6cm}}

\title{AdNanny: One Reasoning LLM for All Offline Ads Recommendation Tasks}

\author{
\makebox[\textwidth][c]{\textbf{Nan Hu, Han Li, Jimeng Sun, Lu Wang, Fangkai Yang, Bo Qiao, Pu Zhao, David Dai,}} \\
\makebox[\textwidth][c]{\textbf{Mengyu Liu, Yuefeng Zhan, Jianjin Zhang, Weihao Han, Allen Sun, Qingwei Lin,}} \\
\makebox[\textwidth][c]{\textbf{Saravan Rajmohan, Dongmei Zhang, Denvy Deng, Feng Sun, Qi Zhang}} \\
\makebox[\textwidth][c]{\{hunan, hanli4, jimengsun, wlu, fangkaiyang, boqiao, puzhao, daviddai, lmengyu, yuefzh,} \\
\makebox[\textwidth][c]{jianjzh, weihan, hasun, qlin, saravar, dongmeiz, dedeng, sunfeng, qizhang\}@microsoft.com}
}

\begin{document}

\maketitle

\pagestyle{fancy}
\thispagestyle{fancy}

\fancyhf{}  

\fancyhead[L]{%
  \vspace{-0.2cm}
  \includegraphics[height=0.9cm]{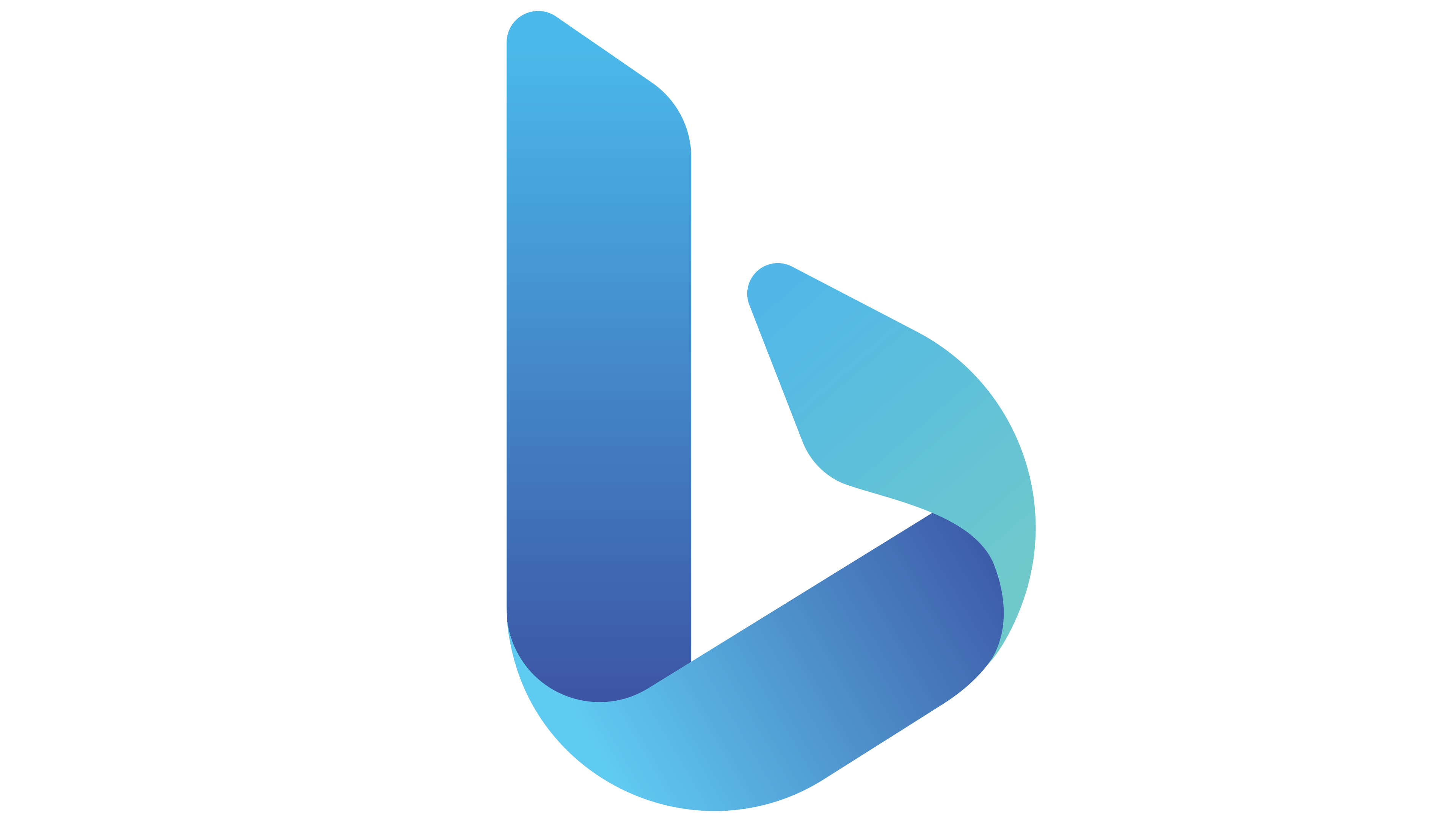}%
}

\fancyhead[C]{}
\fancyhead[R]{}

\fancyfoot[C]{\thepage}

\renewcommand{\headrulewidth}{0.4pt}  

\begin{abstract}
Large Language Models (LLMs) have demonstrated remarkable capabilities across a wide range of Natural Language Understanding (NLU) and Natural Language Generation (NLG) tasks. However, directly deploying LLMs in online advertising systems is often impractical due to strict millisecond-level latency requirements per request, which has led to growing interest in using LLMs \emph{offline} to improve the quality and efficiency of retrieval, ranking, and recommendation models. Existing approaches typically fine-tune separate LLMs for individual tasks (such as query–ad and ad–user relevance labeling, keyword-based query generation, and user profile generation), leading to redundant models, higher maintenance cost, and limited performance despite substantial overlap in domain knowledge and reasoning patterns. In this work, we introduce AdNanny, a single unified LLM that serves as a reasoning-centric backbone for offline ads tasks. AdNanny is built by fine-tuning a public 671B-parameter DeepSeek-R1 checkpoint with a customized Megatron-based trainer that handles its hybrid dense–MoE parallelism, optimizer states, and checkpoint formats, enabling scalable multi-node training while preserving parity with the official implementation. On top of this infrastructure, we construct reasoning-augmented corpora where each example pairs structured supervision with step-by-step natural-language explanations. A multi-task supervised fine-tuning (SFT) stage with adaptive reweighting teaches AdNanny to handle diverse labeling and generation tasks in a consistent label+reasoning format, followed by a reinforcement learning stage that uses downstream ads metrics as feedback to better align its behavior with online retrieval and ranking objectives. AdNanny is deployed in production within Bing Ads, where it has substantially reduced manual labeling effort and improved accuracy across multiple offline tasks. By consolidating numerous task-specific models into one reasoning-centric foundation model, AdNanny delivers a higher-performing, more cost-effective, and scalable solution, offering new insights into how LLMs can be systematically leveraged for large-scale, domain-specific applications.
\end{abstract}

\section{Introduction}
Large language models (LLMs) have achieved state-of-the-art performance on a broad range of natural language understanding (NLU) and natural language generation (NLG) tasks, including question answering, dialogue, summarization, and code-related reasoning~\citep{ourqa,warriorcoder,warriormath,wizardlm}. Their ability to encode rich semantics and world knowledge makes them natural candidates for understanding user queries, ad creatives, and landing pages in modern advertising platforms. However, directly deploying a general-purpose LLM as an end-to-end recommender system in a high-throughput online advertising service such as Bing Ads is currently impractical. LLMs with billions of parameters typically incur inference latencies on the order of hundreds of milliseconds or more, even on specialized accelerators. By contrast, commercial ad serving pipelines must satisfy strict end-to-end budgets of only a few to a few tens of milliseconds for retrieval, ranking and auction logic combined. In addition, general-purpose LLMs often lack specialized knowledge of the advertising domain and require lengthy text inputs to incorporate user behavior, making a naive direct application inefficient. 

To bridge this gap, today’s industrial advertising systems cannot simply replace their existing pipelines with an LLM. Instead, they incorporate LLMs as enhancement modules coupling with industrial ad pipelines in a latency-aware manner along several complementary directions: 

\noindent\textbf{Semantic feature extractor.} A large body of ad systems use LLMs as extractors to enrich item and user representations. Pre-trained or lightly fine-tuned models, such as LEARN framework~\citep{jia2025learn}, NoteLLM-series~\citep{zhang2024notellm, zhang2025notellm}, and MicroBERT used at eBay~\citep{gessler2022microbert, ebay_language_model_2023}, transform titles, descriptions, and reviews into dense embeddings that are concatenated with ID features in the ranking model, significantly mitigating cold-start and long-tail sparsity by allowing new items and users to be modeled purely from text. 

\noindent\textbf{Recall enhancer.} LLMs are employed to augment recall: multi-tower architectures such as RecGPT~\citep{yi2025recgpt} introduce an LLM-driven ``tag tower'' that generates high-level user interest descriptors and matches them with item embeddings to improve semantic coverage and diversity, while generative recall approaches, e.g., GRAM~\citep{pang2025generative}, prompt an LLM with recent behaviors to generate latent intent keywords or pseudo-queries, which are then used in vector retrieval. To accommodate latency constraints, these generations are typically pre-computed offline or distilled into lightweight recall models.

\noindent\textbf{Intent reasoner.} Several platforms leverage LLMs for user intent understanding and scenario reasoning: world-knowledge models, e.g., URM~\citep{jiang2025large}, combine behavior logs with commonsense inference to infer user life stages or contexts (e.g., “home renovation phase,” “parent preparing New Year gifts”) and adjust recommendations accordingly, while other deployments, such as 360 Ads~\citep{zhuanlan-360-llm-ads}, let an LLM synthesize rich, human-readable interest summaries (“25–30-year-old female, currently focused on light-luxury beauty and fitness”) that are fed as dense features into the downstream ranker, improving conversion and faster adaptation to interest drift.

\noindent\textbf{Teacher for knowledge distillation.} To balance effectiveness and efficiency, many works adopt knowledge distillation and compression, where a powerful LLM (or a mid-size domain model like microBERT~\citep{gessler2022microbert}) provides soft labels or intermediate representations used to train much smaller CTR/CVR or recall models that retain most of the accuracy but meet millisecond-level serving budgets~\citep{dey2025llmdistill4ads}.

A broader survey of LLM-based recommenders~\citep{wang2024towards} shows similar design patterns: LLMs are typically used as feature generators, intent taggers, or offline labelers, and their signals are then distilled into smaller task-specific models that satisfy real-time serving constraints. This architecture has clear practical advantages, but it also relies heavily on compact encoders or small/mid-sized LMs that underperform their larger teachers on nuanced language phenomena, especially in long-tail or cross-domain scenarios. At the same time, many offline and near-line tasks in ads (e.g., query–ad relevance judgment, ad–user relevance scoring, creative and keyword generation) share substantial domain knowledge about users, queries, ads, and landing pages, yet are still handled by separate models optimized in isolation. As a result, industrial systems often end up maintaining a zoo of small models: each with its own training data, evaluation protocol, deployment pipeline, and drift issues—despite their overlapping semantics. This fragmentation increases engineering and operational overhead and can lead to inconsistent behavior across components, running counter to the original motivation for LLMs as \emph{unified text interfaces} capable of handling diverse NLU and NLG tasks with a single model. These limitations motivate a different design point: centralizing ads-domain knowledge and supervision in a single, stronger, reasoning-centric LLM that operates offline, and letting latency-critical components in the serving stack consume its label+reasoning outputs or distilled variants rather than training every task-specific model from scratch.

In this work, we take this direction and develop \textbf{AdNanny}, a single, unified LLM that centralizes ads-domain knowledge and supervision while remaining compatible with existing low-latency serving stacks. AdNanny is built by fine-tuning a public 671B-parameter DeepSeek-R1 checkpoint~\citep{liu2024deepseek} on reasoning-augmented corpora constructed from multiple offline tasks, including query--ad relevance labeling, ad--user relevance labeling, keyword-based query generation, etc. Each training example pairs structured supervision (labels or scores) with step-by-step natural-language explanations derived via a unified data-cooking pipeline (Figure~\ref{fig:data_prep}), enabling the model to internalize not only \emph{what} decision should be made, but also \emph{why}. This reasoning-centric design makes AdNanny a single backbone that can consistently support diverse labeling and generation tasks in the ads ecosystem.

To make such a large hybrid dense--MoE model trainable in practice, we build a customized Megatron-based trainer~\citep{shoeybi2019megatron} that handles DeepSeek-R1's parallelism pattern, optimizer states, and checkpoint formats while preserving parity with the official implementation. 
On top of this infrastructure, we construct large-scale reasoning-augmented datasets where each example pairs structured supervision with step-by-step natural-language explanations, and train AdNanny via multi-task supervised fine-tuning with adaptive task-level and instance-level reweighting to balance heterogeneous offline workloads. A second-stage reinforcement learning (RL)~\citep{ouyang2025token,wang2025text2grad,wang2025lettingo} phase then aligns the model directly with downstream ads metrics such as retrieval, ranking, and recommendation quality, using offline evaluators as rewards. 
Deployed as an internal offline labeling and generation service within Bing Ads, AdNanny has substantially reduced manual annotation effort, improved accuracy across multiple offline tasks, and allowed us to retire a collection of task-specific small models in favor of a single reasoning-centric backbone whose outputs can be directly consumed or distilled by downstream components.

\section{Offline tasks}\label{sec:offline-tasks}
The Bing Ads serving ecosystem depends on a wide range of offline tasks that create the labels, testbeds, and representations used to train and monitor online models. These tasks do not directly serve ads to users; instead, they shape the supervision signals and evaluation datasets that online retrieval, ranking, and creative systems rely on. AdNanny is designed as an offline assistant for this ecosystem, unifying and automating five key categories of tasks and providing consistent, reasoning-based outputs that downstream components can consume. 

\noindent\textbf{Offline query–ad and ad–user relevance labeling.} Offline relevance labeling aims to judge whether an ad is semantically and contextually relevant to a given query or user intent, often with graded relevance levels (e.g., bad / fair / good ). These labels are used to train and calibrate lightweight relevance models, to construct offline evaluation sets for retrieval and ranking, and to annotate samples from online A/B tests. They complement behavioral signals such as CTR, which reflect user behavior under a specific ranking policy and can be noisy with respect to true semantic relevance. AdNanny generates fine-grained, reasoning-augmented relevance judgments that explicitly explain why an ad is relevant or irrelevant. Its labels can train or refresh small relevance models, benchmark candidate online models on relevance quality, and help practitioners debug failure cases with reasoning context, thereby improving both coverage and transparency without relying solely on expensive human labeling.

\noindent\textbf{Model evaluation and replacement analysis.} Model evaluation and replacement analysis compare a candidate model against a production baseline on quality, safety, and policy adherence before the candidate is exposed to full online traffic. Offline evaluation serves as a gatekeeper: it filters out clearly underperforming variants, identifies systematic risks (e.g., policy-violating creatives or low-relevance matches), and selects a small number of promising candidates for live A/B testing, thereby reducing the cost and risk of experimentation. AdNanny acts as a reasoning-based evaluator that inspects model outputs on shared prompts or sampled traffic, produces natural-language comparisons of strengths, weaknesses, and potential risks, and aggregates these into offline signals or reports that guide which models should proceed to online A/B tests, reducing manual judging effort and speeding up iteration.

\noindent\textbf{Query generation.} 
Query generation leverages an ad’s creatives, landing page, and metadata to synthesize search queries that are relevant to that ad or ad group and could realistically be typed by users. In the Bing Ads ecosystem, this task makes explicit the ``query surface'' under which an ad should be discoverable, especially in long-tail and cold-start scenarios.
Given ad-side content (titles, descriptions, landing page snippets, and so on), AdNanny generates plausible user queries together with short explanations of which parts of the ad they correspond to. The resulting query–ad pairs can be used as labeled training examples, as offline test suites for recall evaluation, or as seeds for further filtering and deduplication, making AdNanny a controllable, interpretable query generator rather than a black-box data augmenter. 
Beyond ad-side synthesis, query generation can also be user-conditioned, where the model derives intent-representative queries from a user’s recent interaction history (e.g., clicked ads, browsed pages, and engagement signals). These user-conditioned queries act as a complementary for ads recall, supplementing the explicit user-issued query to retrieve additional relevant ad candidates. This is particularly useful when the current query is underspecified or ambiguous, enabling improved coverage for long-tail interests and multi-step exploratory search behavior.

\noindent\textbf{User profile generation.}
User profile generation summarizes a user’s historical behavior into structured descriptions of their interests, intents, and constraints, e.g., preferred categories or topics that should be avoided, in a form that models or analysts can directly consume. These profiles are used to build features for personalization in retrieval, ranking, and CTR prediction, and to define user segments for offline analysis and evaluation, for example, ``users interested in travel but not luxury goods'', so that models are tuned and assessed beyond simple global averages. From logs of queries, impressions, clicks, and conversions, AdNanny produces textual profiles describing likely preferences, intent patterns, and risk-sensitive attributes; these profiles can then be encoded into embeddings or discrete features for model training, or inspected directly in human-readable form to support offline diagnostics and debugging.

\noindent\textbf{Ad optimization and keyword extension.}
Ad optimization and keyword extension aim to improve how ads are written and where they can be matched by rewriting creatives and expanding keyword lists with semantically related variants. Offline, these operations help advertisers and systems explore more diverse ad texts, improve writing quality, and uncover additional queries under which an ad should reasonably be shown, thereby increasing both coverage and candidate diversity. The resulting rewrites and keyword candidates are reviewed, filtered, and then passed to online systems for recall and creative optimization. AdNanny suggests alternative ad texts, titles, and keywords with short rationales, for example, ``same puirchase intent but emphasizes price'', and they can be integrated into internal tools for human review or converted into training pairs for creative-ranking and keyword-suggestion models, enabling a continuous offline improvement loop while keeping the online serving path lightweight.

Taken together, these offline tasks collectively cover the critical reasoning, evaluation, and generation capabilities required for unified offline ads intelligence. AdNanny provides a single LLM-based interface that can label relevance, critique and compare models, generate queries, summarize users, and optimize ads, while exposing consistent and interpretable outputs to downstream systems. In the next section, we describe how we construct datasets for these tasks and how we prepare the training data that enables AdNanny to serve as a general-purpose offline assistant for Bing Ads.

\section{Data Preparation}
\label{sec:data}

The performance of AdNanny hinges on the quality and reasoning richness of its training data.
Conventional ads datasets typically provide only discrete supervision (e.g., a relevance label or a click signal), which is sufficient for training black-box predictors but insufficient for teaching a model \emph{how} to reason and justify its decisions.
By contrast, AdNanny relies on \emph{reasoning-augmented supervision}: each training instance pairs structured labels with a step-by-step natural-language explanation of the underlying decision logic.
This not only enhances interpretability, but also enables the model to perform human-like evaluation and to support multiple offline tasks (Section~\ref{sec:offline-tasks}) with a unified reasoning engine.

Although we previously introduced five offline tasks, their data preparation follows a shared pattern.
In this section, we describe this pattern using three representative tasks as running examples:
(1) \textbf{query–ad relevance labeling} (query–ad pairs with labels),
(2) \textbf{ad-user relevance labeling} (user behavior plus ad with a relevance score), and
(3) \textbf{keyword-based query generation} (generating user queries from advertiser keywords and ads).
The remaining tasks reuse the same ``cooked'' reasoning-augmented pipeline.

\subsection{Unified reasoning-augmented pipeline}

Across these three representative tasks, we start from existing supervised or rule-constrained data:

\begin{itemize}
\item A \emph{query–ad relevance} dataset provides triples ⟨query, ad, labels⟩, where the labels may include overall relevance, location relevance, and ad-quality judgments.
\item A \emph{user history–ad relevance} dataset provides ⟨user history, ad, label⟩ examples, where the label is an ordinal relevance score measuring how well the ad matches the user’s interests.
\item A \emph{keyword-based query} dataset provides ⟨keyword, generated query⟩, where the queries are generated and optimized from keywords following pre-defined matching rules.
\end{itemize}

In all three cases, the raw data already encodes ``what decision was made'' (the label or rule), but not \emph{why}.
The goal of data preparation is therefore to convert these into reasoning-augmented examples such as ⟨query, ad, labels, reasoning⟩, ⟨user history, ad, label, reasoning⟩, or ⟨keyword, generated query, reasoning⟩.

\begin{figure}[!t]
\centering
\includegraphics[width=0.95\linewidth]{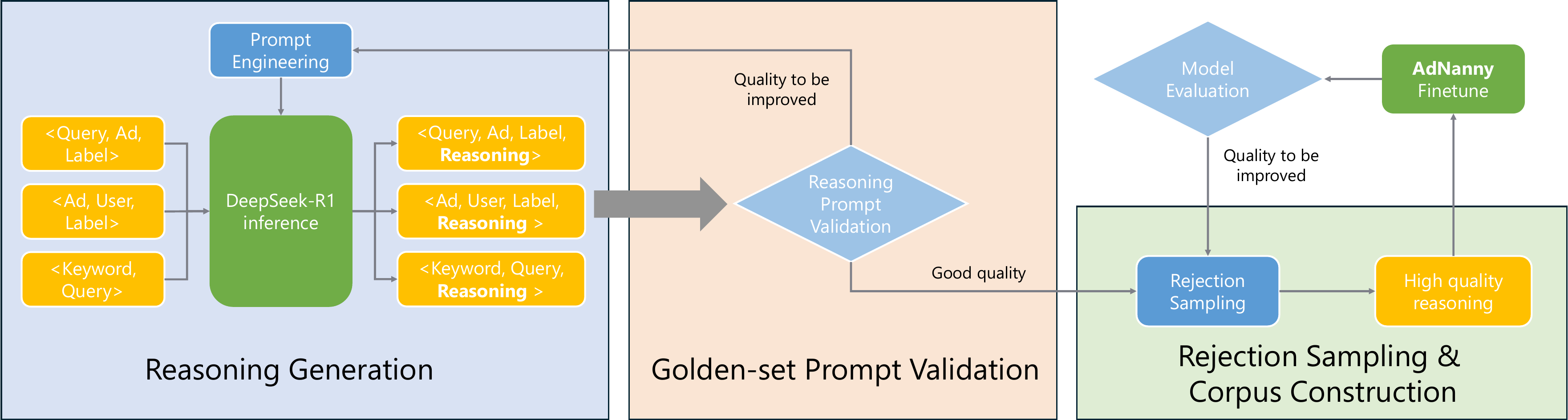}
\caption{Unified reasoning-augmented data preparation pipeline.
DeepSeek-R1 first uses task-specific \emph{reasoning prompts} to generate label+reasoning candidates (\emph{Reasoning Generation}); these reasoning prompts are then validated and refined on golden sets (\emph{Golden-set Prompt Validation}), before being applied at scale with rejection sampling to construct the high-quality corpus used to fine-tune AdNanny.}
\label{fig:data_prep}
\end{figure}

To do this, as shown in Figure~\ref{fig:data_prep}, we apply a unified pipeline with three stages: \textbf{reasoning generation}, where DeepSeek-R1 is prompted with task-specific instructions that take existing labels or rule-constrained outputs as input and produce step-by-step explanations; \textbf{golden-set prompt validation}, where these prompts are tested and refined on small curated golden sets until the generated reasoning is faithful, well formatted, and robust; and \textbf{rejection sampling and corpus construction}, where the validated prompts are run at scale, low-quality outputs are filtered using schema and self-consistency checks, and the retained reasoning-augmented examples are standardized, shuffled, and split into training and evaluation sets for fine-tuning the DeepSeek models that power AdNanny.

\subsection{Reasoning generation for representative tasks}\label{sec:reasoning_geneartion}

For the query–ad and ad–user relevance tasks, the inputs to the ``data cooking'' prompts are already-labeled examples.
The prompts do not ask the model to choose a new label; instead, they give the existing label and ask the model to reconstruct the reasoning that makes this label appropriate.

For \emph{query–ad relevance labeling}, each instance consists of a query, an ad (title, description, URL), and a set of categorical labels, including overall relevance, location relevance, and ad-quality.
The prompt to DeepSeek-R1 provides all of these and then asks the model, in a few explicit steps, to describe the user’s intent, summarize what the ad offers, and explain why the given labels (e.g., bad / fair / good) make sense. A lightweight example is:

\begin{tcolorbox}[colback=gray!5,colframe=gray!20]
\small
\textbf{Query:} cheap hotel in downtown Paris\\ 
\textbf{Ad title:} Luxury 5-star hotel in central Paris\\
\textbf{Ad description:} Premium suites, fine dining, and concierge service near major attractions \dots\\
\textbf{Ad URL:} \texttt{www.examplehotel.com}\\[2pt
]
\textbf{Location label:} \texttt{Good}\\
\textbf{Relevance label:} \texttt{Fair}\\
\textbf{Ad-quality label:} \texttt{Good}
\end{tcolorbox}

For this example, the model would produce a \texttt{<think>} block such as:
\emph{``The user wants an inexpensive place to stay in downtown Paris.
The ad is for a luxury 5-star hotel in central Paris.
The city and central location match the user’s intent, but the price level does not match the request for something cheap, so the ad only partially satisfies the need.
The ad text itself is clear and well written.
Therefore the location label `Good', relevance label `Fair', and ad-quality label `Good' are appropriate''.}
It then repeats these labels in \texttt{<LocationLabel>}, \texttt{<Label>}, and \texttt{<AdQuality>} tags.
This transforms the original instance into a reasoning-augmented example ⟨query, ad, label, reasoning⟩.

For \emph{ad–user relevance labeling}, the input is a sequence of recent user interactions (queries, page views, clicks, purchases), an ad, and an existing relevance score (e.g., from 0 to 3).
The prompt follows the same pattern as for query–ad relevance: it provides the user history, ad, and score, and asks the model to summarize the user’s main interests and to explain why the given score is appropriate. For example:

\begin{tcolorbox}[colback=gray!5,colframe=gray!20]
\small
\textbf{User history:} searched `beginner 5k training plan'; clicked `how to choose running shoes'; viewed several product pages for running shoes and socks\\
\textbf{Ad title:} Lightweight running shoes for new runners\\
\textbf{Ad description:} Comfortable cushioned shoes designed for beginner joggers \dots\\
\textbf{Relevance score:} \texttt{3}
\end{tcolorbox}

For this example, the model may write a \texttt{<think>} block such as:
\emph{``The user’s recent activity is all about starting to run and researching running shoes and related gear.
The ad promotes lightweight, cushioned running shoes specifically aimed at new runners, which directly matches these interests.
Therefore a relevance score of 3 (highly relevant) is appropriate''.}
This yields a reasoning-augmented tuple ⟨user history, ad, score, reasoning⟩, which can support both user profiling and user–ad relevance tasks.

The \emph{keyword-based query generation} task fits into the same pipeline, but has an additional generation step before reasoning is attached.
We start from advertiser keywords, associated ads and landing pages, and a set of matching and quality rules. For example, generated queries must stay in the same product category as the keyword, be \emph{narrower} and more specific, and follow strategic principles such as using natural synonyms only when they keep strong relevance, adding domain-specific terms, incorporating brand or purpose-driven context, varying user intents (e.g., buying vs.\ researching), and adding realistic constraints like budget or ``near me''. Using a first DeepSeek-based generation prompt, we create and optimize queries that follow these rules, yielding an initial dataset of ⟨keyword, generated query⟩ pairs.

\begin{tcolorbox}[colback=gray!5,colframe=gray!20]
\small
\textbf{Keyword:} laptop computer\\[2pt] \textbf{Optimized queries (following the rules):}\\ \hspace*{1em}-- ``gaming notebook under 1000'' (synonym + price context, maintains relevance)\\
\hspace*{1em}-- ``Dell laptop reviews 2024'' (brand + evaluation intent, highly relevant)\\ \hspace*{1em}-- ``business computer portable'' (work context + wording variation, relevant)\\
\hspace*{1em}-- ``refurbished laptops for students'' (condition + target user, relevant)
\end{tcolorbox}

Given this dataset, a second ``reasoning'' prompt is used for data cooking: it shows the keyword, one of the generated queries, and a short reminder of the rules, and asks DeepSeek-R1 to briefly explain why this query is a suitable, rule-compliant refinement of the keyword. For example, for the pair \(\langle\)``laptop computer'', ``Dell laptop reviews 2024''\(\rangle\), the model may write: \emph{The query keeps the laptop-computer category, adds a specific brand (Dell), and introduces an evaluation intent (reviews in 2024), making it narrower and still highly relevant to users interested in buying a laptop.''} Attaching such explanations gives us reasoning-augmented examples ⟨keyword, generated query, reasoning⟩ that can later be used for training the query generation and ad optimization components of AdNanny.

\subsection{Golden-set prompt validation}

To ensure that the generated reasoning is reliable and aligned with task definitions, we first perform \emph{golden-set} validation for the reasoning prompts. For each of the three representative tasks, we construct a small golden set of high-quality examples: query–ad pairs with trusted labels, user history–ad pairs with reliable scores, and keyword–query pairs that satisfy the matching rules. We run DeepSeek-R1 with the corresponding reasoning prompts on these golden examples and manually inspect whether the reasoning is faithful to the input, whether it correctly justifies the given labels or rules, and whether the outputs adhere to the required format (for example, \texttt{<think>} and \texttt{<Label>} or \texttt{<Score>} tags are present and well-formed).

To stress-test robustness, we allow mild prompt variations on the golden set (e.g., rephrasing questions such as ``Why might this ad be clicked?'' vs.\ ``What makes this ad relevant?'' or changing the order of reasoning steps) and check that the underlying logic remains consistent. If systematic issues are observed, such as misinterpreting relevance levels, ignoring important parts of the user history, or loosely applying the keyword-matching guidelines, we refine the reasoning prompts by clarifying instructions, adding small in-prompt examples, and tightening constraints, then re-evaluate on the same golden sets. Only prompts that behave robustly under these checks are used in the subsequent large-scale data generation stage.

\subsection{Rejection sampling and corpus construction}

Once a reasoning prompt has passed golden-set validation, we apply it at scale and perform rejection sampling to construct high-quality reasoning-augmented corpora. For each input instance, DeepSeek-R1 may generate multiple reasoning candidates under slight prompt or decoding variations; we then apply a light-weight \emph{self-consistency} filter that measures semantic agreement between candidates (using embedding similarity or entailment-style scoring) and discards clearly divergent or hallucinated explanations. Each surviving reasoning trace is further evaluated by a smaller verifier model, which checks textual alignment with the input, logical soundness, and adherence to task-specific guidelines. In practice, we focus on three aspects: \emph{faithfulness} (the reasoning must support the provided label and avoid contradictions), \emph{consistency} (reasoning for similar inputs should not conflict), and \emph{diversity} (different but valid perspectives, such as user-intent factors vs.\ lexical overlap, are encouraged rather than collapsed into a single template).

Across all offline tasks, we discard outputs that violate the schema (for example, missing required tags or malformed blocks), contradict the existing supervision in labeling tasks, or degenerate into low-information boilerplate that does not reference concrete details in the input. For keyword-based query data, we additionally filter out \(\langle\)keyword, generated query\(\rangle\) pairs whose queries are broader than the keyword, off-topic with respect to the ad and landing page, or near-duplicates of other queries for the same keyword. Periodic manual spot-checks are used to refine these filters and keep the retained reasoning traces both faithful and useful.

After this multi-stage rejection process, we obtain unified reasoning-augmented corpora for the three representative tasks: query–ad relevance (\(\langle\)query, ad, labels, reasoning\(\rangle\)), ad–user relevance (\(\langle\)user history, ad, relevance score, reasoning\(\rangle\)), and keyword-based query generation (\(\langle\)keyword, generated query, reasoning\(\rangle\)). These corpora are then standardized, shuffled, and split into training and evaluation sets, and used to fine-tune the DeepSeek models that serve as AdNanny’s backbone. Although we describe only three tasks explicitly, the same ``cook with reasoning'' pipeline is applied to the remaining offline tasks, providing a common reasoning substrate that supports relevance labeling, model evaluation, query generation, user profiling, and ad optimization within a single unified framework.

\section{Training Pipeline}

AdNanny is obtained by fine-tuning a public DeepSeek-R1 checkpoint on the reasoning-augmented corpora described in Section~\ref{sec:data}. Because DeepSeek-V3/R1~\citep{liu2024deepseek} is a 671B-parameter hybrid dense–MoE model with 37B activated parameters per token and consumes over 15TB memory in full precision, we build a customized Megatron-based trainer~\citep{shoeybi2019megatron} that can efficiently handle its parallelism pattern, optimizer state, and checkpoint formats while preserving parity with the official HuggingFace implementation.
On top of this infrastructure, AdNanny is trained in two stages: first, a multi-task supervised fine-tuning (SFT) stage where the model learns to generate label+reasoning targets for all offline tasks under a unified autoregressive objective with adaptive instance- and task-level reweighting; and second, a reinforcement-learning stage where the SFT model is further aligned with downstream ads metrics via a KL-regularized GRPO-style objective over both classification-style and generation-style tasks.

\subsection{Parallelism and Efficiency}
\subsubsection{Megatron-based parallelism design}\label{sec:training}

We adopt Megatron as the core training framework to support DeepSeek’s combination of dense and MoE layers with advanced parallelism. While HuggingFace+DeepSpeed~\citep{deepspeed} is convenient for small-scale experimentation, it only partially supports DeepSeek’s expert, tensor, and pipeline parallelism, which limits throughput on large GPU clusters. In contrast, Megatron exposes flexible pipeline, data, tensor, and expert parallelism, making it a better fit for large-scale AdNanny fine-tuning.

\begin{figure}[!t]
\centering
\includegraphics[width=0.7\linewidth]{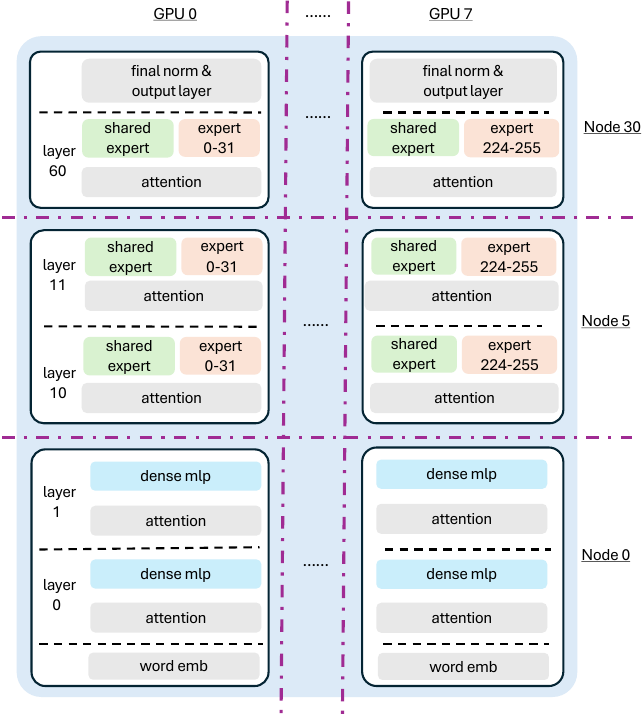}
\caption{DeepSeek-R1 Parallelism Design.}
\label{fig:para_pipe}
\end{figure}

Our deployment uses 248 GPUs across 31 nodes  and follows two partitioning principles: (i) distribute parameters and FLOPs as evenly as possible across devices, and (ii) minimize cross-node communication, especially for MoE traffic. With regards to this, we design a hybrid parallel configuration (shown in Figure~\ref{fig:para_pipe}):

\begin{itemize}
\item \emph{Pipeline parallelism (31-way PP).} The model is sliced into 31 stages, assigning roughly 1–2 transformer layers per node from the word embedding and early dense layers to the final normalization and output head. This allows us to fit the full DeepSeek-R1 model while keeping the activation footprint per stage manageable.
\item \emph{Expert parallelism (8-way EP) for MoE layers.} Each MoE block has a \emph{shared expert} that every token always visits, plus 256 routed experts. In our EP layout, the routed experts are split across 8 GPUs (each GPU holds a disjoint subset such as experts 0–31 or 224–255), and the shared expert is replicated on every EP rank. This keeps expert communication localized within an 8-GPU node group and lets tokens use the shared path without cross-node traffic.
\item \emph{Data parallelism (8-way DP) for dense and shared components.} Non-MoE layers and shared experts are replicated across 8 data-parallel groups. Gradients are synchronized only for these shared parameters, avoiding unnecessary communication for purely local experts.
\end{itemize}

This layout balances memory and compute across nodes and avoids hotspots on specific stages. The same partition is used in both pre-fine-tuning warm-up and AdNanny multi-task training, so that model parallel states and optimizer states remain consistent throughout the pipeline.

\begin{figure}[!t]
\centering
\includegraphics[width=0.95\linewidth]{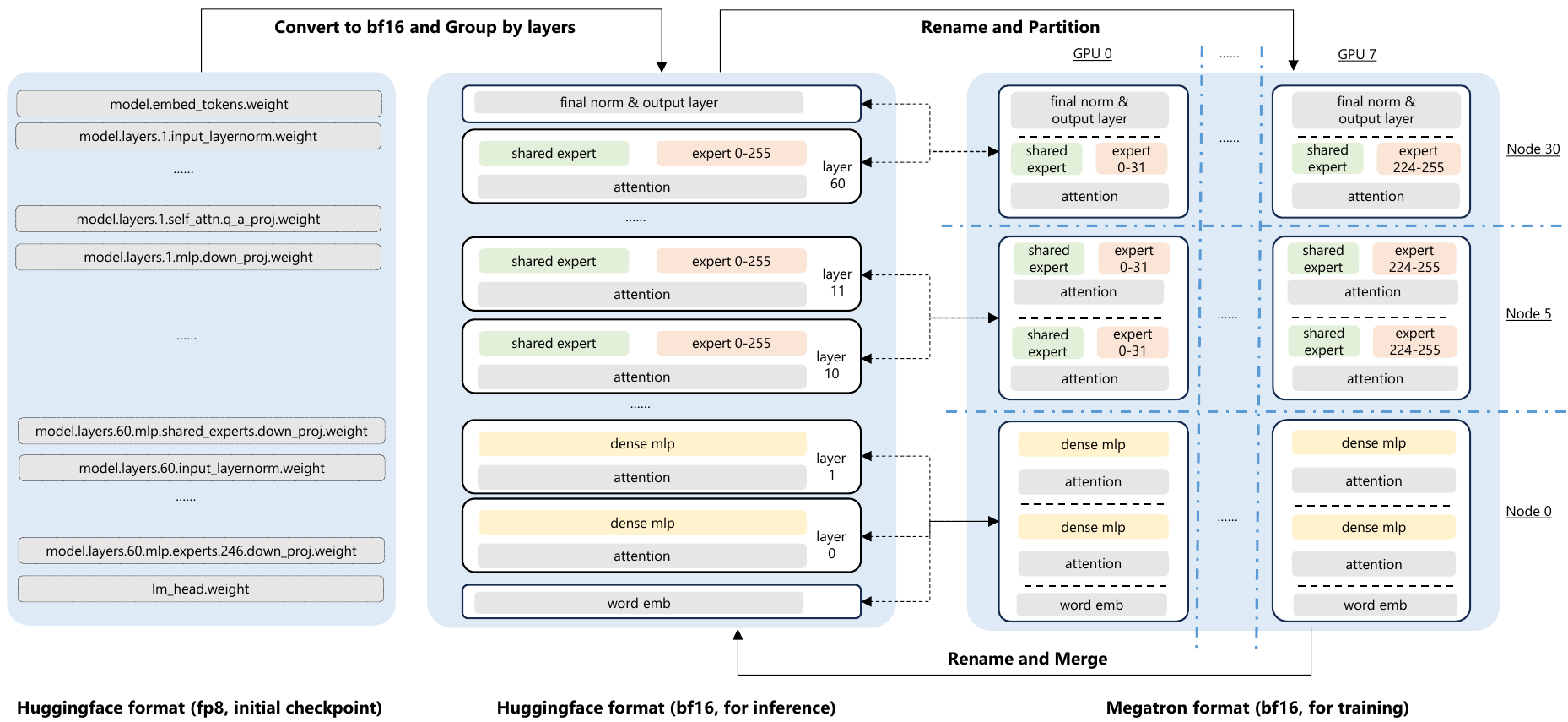}
\caption{Checkpoint conversion and sharding.}
\label{fig:checkpoint}
\end{figure}

\subsubsection{Checkpoint conversion and sharding}

The public DeepSeek-R1/V3 checkpoints are released in a HuggingFace-style format, typically stored in FP8 for efficiency and organized as a flat dictionary of tensors with string keys such as \texttt{model.layers.0.self\_attn.q\_proj.weight} or \texttt{model.layers.1.mlp.gate\_proj.weight}. This format is convenient for single-node inference, but it is not directly compatible with our Megatron-based, multi-parallel training setup. 
To increase throughput, we integrate optimized fused kernels from TransformerEngine~\citep{nvidia_transformer_engine_2022}, replacing baseline attention, normalization, and MoE routing operators. Training runs in mixed precision with \textit{BF16 computation} and \textit{FP32 accumulation}. Several dense and MoE kernels are re-implemented with explicit FP32 accumulation during partial-sum stages, preventing numerical overflow while preserving training stability.

We therefore implement a bidirectional conversion pipeline (shown in Figure~\ref{fig:checkpoint}) that (i) transforms the distribution checkpoint into a training-ready, sharded Megatron checkpoint, and (ii) can merge the trained shards back into a standard HuggingFace checkpoint for downstream serving. The forward direction proceeds in two main steps:

\begin{enumerate}
\item \emph{Convert to \textbf{BF16} and group by layers.} 
The FP8 tensors are loaded and cast to bf16, which is our standard training precision. We then parse the parameter names and reconstruct explicit layer objects: for each key containing \texttt{model.layers.$i$.}, we extract the layer index $i$, and move that tensor into the corresponding layer container. Conceptually, this step turns a flat list of thousands of tensors into an ordered list of transformer blocks (e.g., Layer 0, Layer 1, $\dots$), where each block now aggregates its attention, MLP, and (when applicable) MoE expert weights. This ``group by layers'' phase is what makes structured sharding possible: once we know exactly which weights belong to each depth, we can slice them consistently across GPUs.
\item \emph{Rename and shard.} 
These layer-wise weights according to the parallel layout described in Section~\ref{sec:training}. For MoE layers, the 256 experts (indexed 0–255) are partitioned into contiguous ranges and assigned to specific expert-parallel ranks (for example, a given GPU may host the shared expert plus experts 224–255). The script rewrites parameter names into Megatron’s expected naming scheme and emits one checkpoint shard per rank, containing exactly the parameters that rank is responsible for during training. The reverse direction follows the opposite path: Megatron shards are loaded, merged by layer and expert index, and written back into a single HuggingFace checkpoint in FP8 or BF16, enabling unified inference or further distribution through standard tooling.
\end{enumerate}

The result is a Megatron-compatible BF16 checkpoint that can be loaded directly on the 248-GPU cluster for AdNanny fine-tuning, and later converted back to HuggingFace format for inference and deployment.

\subsubsection{Distributed optimizer for hybrid dense–MoE layers}

DeepSeek-R1 mixes dense blocks and MoE layers, and Megatron’s default distributed optimizer maintains \emph{separate} optimizer groups for dense and MoE parameters. In our parallel layout, the first pipeline stage (Node~0) hosts only embeddings and early dense layers, while MoE layers live on Nodes~1–30. This creates an asymmetry: during each training step, all ranks participate in dense-gradient collectives, but only Nodes~1–30 have MoE parameters and enter the MoE collectives. Node~0 skips these MoE synchronizations entirely, so the other ranks block inside the MoE all-reduce waiting for a participant that never arrives, causing training to hang.

We resolve this by instantiating a lightweight \emph{stub MoE optimizer} on Node~0. The stub owns no real parameters but joins the same MoE process group as the other ranks and participates in all MoE collectives with empty tensors. This restores a uniform communication pattern where every rank now calls dense and MoE synchronizations in the same order, eliminating deadlocks while preserving the performance benefits of separate dense and MoE optimizers on nodes that actually host MoE layers.

\subsubsection{Trainer verification}

Given the complexity of the model and training stack, we first verify that our Megatron-based trainer is faithful to the reference DeepSeek implementation before applying it to AdNanny.

\noindent\textbf{Forward-prop parity.}
We align numerical behavior at the operator level with the HuggingFace DeepSeek-V3/R1 implementation: embeddings, attention, MoE gating, normalization, and output heads use the same precision settings and key hyperparameters (e.g., rotary scaling factors and layer-norm epsilon). We then compare end-to-end performance on public benchmarks. On GSM8K~\citep{cobbe2021training}, HumanEval~\citep{chen2021evaluating}, and MMLU~\citep{hendrycks2020measuring}, the Megatron trainer produces checkpoints whose accuracies differ from the HuggingFace baselines by less than one percentage point (GSM8K: 90.49\% vs.\ 90.37\%, HumanEval: 86.62\% vs.\ 87.26\%, MMLU: 88.67\% vs.\ 88.43\%), confirming functional parity.

\noindent\textbf{Training behavior.}
We further validate training behavior on an internal Extended Phrase Match (EPM) task used in our ads stack with Mistral~\citep{jiang2023mistral7b}, where the Megatron trainer matches the HuggingFace baseline in both loss and final AUC (96.92 vs.\ 96.90), indicating that the data pipeline, optimizer, and schedules are correctly configured. For DeepSeek-R1, Megatron training likewise shows stable, monotonically decreasing loss with no divergence, matching the qualitative behavior reported in the original DeepSeek recipes.

These checks give us confidence that our parallelism layout, checkpoint conversion, and distributed optimizer do not introduce hidden regressions, and that the resulting AdNanny model can be treated as a drop-in, production-ready variant of DeepSeek-R1 for offline Ads tasks.

\subsection{Multi-Task Supervised Fine-Tuning}

AdNanny is trained with multi-task supervised fine-tuning over the reasoning-augmented datasets described in Section~\ref{sec:data}. Each offline task \(t\) (e.g., query–ad relevance, ad–user relevance, keyword-based query generation) provides examples of the form
\(\langle x_{t,i}, y_{t,i} \rangle\), where for each sample \(i\), \(x_{t,i}\) is the input (query, ad, user history, keyword, etc.) and \(y_{t,i}\) is a structured textual target that includes both the discrete label(s) and the natural-language reasoning (e.g., \texttt{<Label>Fair</Label>} plus a \texttt{<think>} explanation).

\subsubsection{Multi-task objective with reasoning-augmented targets}

For a sample \(i\) from task \(t\), we treat the target \(y_{t,i}\) as a single output sequence of tokens \((y_{t,i,1}, \dots, y_{t,i,K_{t,i}})\) that concatenates label tokens and reasoning tokens. Given input \(x_{t,i}\), the supervised loss is the standard autoregressive cross-entropy over the full target:
\begin{equation}
\ell_{t,i}(\theta)
=
- \sum_{k=1}^{K_{t,i}}
\log p_\theta\bigl(y_{t,i,k} \mid x_{t,i}, y_{t,i,<k}\bigr),
\end{equation}
where \(\theta\) denotes the AdNanny parameters. This encourages the model to both predict the correct labels and reproduce a coherent, task-aligned reasoning trace.

Across tasks, we adopt a weighted multi-task objective:
\begin{equation}
\mathcal{L}(\theta)
=
\sum_{t} \lambda_t
\, \mathbb{E}_{i \sim \mathcal{D}_t}
\bigl[\, w_{t,i} \, \ell_{t,i}(\theta) \,\bigr],
\end{equation}
where \(\mathcal{D}_t\) is the dataset for task \(t\), \(w_{t,i}\) is an instance-level weight, and \(\lambda_t\) is a task-level weight. Both \(w_{t,i}\) and \(\lambda_t\) are adjusted during training to focus learning on underfit data and underperforming tasks.

\subsubsection{Instance-level reweighting via perplexity progress}

Within each task, we adaptively reweight individual examples based on how much the model has improved on them compared with a previous checkpoint. Intuitively, if the model has already learned to explain and label a sample well, its perplexity on that sample will drop significantly; if the model has not learned much, the perplexity will change little, suggesting that this region of the data deserves more attention.

Let \(\theta^{(k-1)}\) and \(\theta^{(k)}\) denote the model parameters at two consecutive checkpoints (e.g., at the end of epochs \(k-1\) and \(k\)). For a sample \(i\) in task \(t\), we define its per-sample perplexity under checkpoint \(\theta\) as
\begin{equation}
\mathrm{ppl}_{t,i}(\theta)
=
\exp\!\Bigl(
\frac{1}{K_{t,i}}
\sum_{k=1}^{K_{t,i}}
- \log p_\theta\bigl(y_{t,i,k} \mid x_{t,i}, y_{t,i,<k}\bigr)
\Bigr).
\end{equation}
We then measure the \emph{learning progress} on that sample as
\begin{equation}
\Delta_{t,i}
=
\mathrm{ppl}_{t,i}(\theta^{(k-1)})
-
\mathrm{ppl}_{t,i}(\theta^{(k)}).
\end{equation}

Large positive \(\Delta_{t,i}\) means the model's perplexity has dropped substantially (the sample is being learned well); small or negative \(\Delta_{t,i}\) indicates little progress or even degradation. We translate this into an instance weight
\begin{equation}
w_{t,i}
=
\mathrm{clip}\Bigl(
\exp(-\beta \, \Delta_{t,i}),
\, w_{\min}, \, w_{\max}
\Bigr),
\end{equation}
where \(\beta > 0\) controls the sensitivity to progress and \(\mathrm{clip}(\cdot, w_{\min}, w_{\max})\) bounds the weights for stability. With this form, samples with large \(\Delta_{t,i}\) (already well learned) receive smaller weights, while samples with small or negative \(\Delta_{t,i}\) are upweighted. This creates an instance-level curriculum that shifts focus toward examples whose reasoning and labels the model has not yet mastered.

In practice, we approximate \(\Delta_{t,i}\) using mini-batch estimates, rather than computing exact perplexity for every sample at every step, to keep the overhead manageable.

\subsubsection{Task-level adaptive weighting}

In addition to per-sample reweighting, we adapt the task-level weights \(\lambda_t\) based on each task's validation performance. The goal is to prevent the model from over-optimizing tasks that are already strong while neglecting tasks where AdNanny is still weak.

Let \(M_t^{(k)} \in [0,1]\) denote a normalized validation metric for task \(t\) at checkpoint \(k\) (e.g., a task-specific score). We define a difficulty signal
\begin{equation}
d_t^{(k)} = 1 - M_t^{(k)},
\end{equation}
so that high-scoring tasks have small \(d_t^{(k)}\) and low-scoring tasks have large \(d_t^{(k)}\). We then set the task weight for the next training interval as
\begin{equation}
\lambda_t^{(k+1)}
=
\frac{\bigl(d_t^{(k)}\bigr)^\alpha}{\sum_{t'} \bigl(d_{t'}^{(k)}\bigr)^\alpha},
\end{equation}
with \(\alpha \ge 1\) controlling how aggressively we emphasize underperforming tasks. When a task's validation metric improves, its \(d_t^{(k)}\) shrinks and its relative sampling weight \(\lambda_t^{(k+1)}\) decreases; tasks with weaker validation performance receive larger weights. This yields a simple, interpretable mechanism for redistributing training capacity toward tasks where AdNanny still has room to improve.

Combining instance-level weights \(w_{t,i}\) with task-level weights \(\lambda_t\), the multi-task SFT procedure continuously steers learning toward both underfit examples and underperforming tasks, while still using the full reasoning-augmented targets for all offline Ads tasks.

\subsection{Reinforcement learning with downstream ads rewards}

While multi-task SFT teaches AdNanny to follow offline label+reasoning supervision, our end goal is to improve the performance of downstream online ads models (retrieval, ranking, recommendation, and other consumers of AdNanny’s outputs). To align AdNanny with these online objectives, we add a second alignment stage that treats AdNanny as a policy and uses the performance of downstream models trained on or driven by its outputs, as an on-policy reward signal~\citep{DBLP:conf/icml/YuanPCLSXW24,wang2025lettingo,DBLP:conf/iclr/HuangFWY0LL0R025}.

\subsubsection{Rewards from downstream ads models}
\label{sec:rl-reward}

We reuse the task index \(t \in \mathcal{T}\) and datasets \(\mathcal{D}_t\) introduced in the SFT objective: each offline task defines a distribution over inputs \(x \sim \mathcal{D}_t\).
For each sampled \(x\), AdNanny with parameters \(\theta\) produces an output
\(y \sim \pi_\theta(\cdot \mid x)\), such as a label+reasoning tuple or a set of generated queries.
A collection of downstream ads models then consume these outputs and define rewards.

Let \(f_t(\cdot; \phi_t)\) denote the downstream model associated with task \(t\), with parameters \(\phi_t\), and let \(M_t(\phi_t; \mathcal{V}_t)\) be a scalar metric (e.g., AUC, NDCG, or a business-weighted objective) evaluated on a fixed validation set \(\mathcal{V}_t\).

For such tasks, we keep downstream
models fixed and use their scoring behaviors as reward providers. Concretely, for each task $t$, AdNanny generates a batch $B_t = \{(x_j, y_j)\}$ of
queries, profiles, or creatives. These outputs are injected into the existing
retrieval and ranking pipelines without retraining any downstream model. We then
evaluate their incremental effect using standard offline metrics such as recall@K,
NDCG@K, and predicted CTR uplift measured by the frozen ranking/CTR models. Let
$M_t^{\text{base}}$ denote the metric of the baseline pipeline, and
$M_t^{\text{with}}$ the metric after incorporating AdNanny’s outputs. The batch
reward is defined as:

\begin{equation}
R_t(B_t) = g_t\!\Bigl( M_t^{\text{with}} - M_t^{\text{base}} \Bigr),
\end{equation}

and we assign $r_t(x_j, y_j) = R_t(B_t)$ to all samples in $B_t$. In practice, we
primarily use recall@K for retrieval-oriented generation and predicted CTR uplift
from frozen CTR models for ranking-oriented generation. This directly aligns
AdNanny’s generations with downstream online utility, while remaining fully
compatible with production constraints.

\section{Experiments}

\subsection{Query--Ad Relevance Evaluation}
\label{sec:exp-qac}

We first evaluate AdNanny on internal \emph{query--ad relevance} tasks derived from production judgments. Refer to Section~\ref{sec:reasoning_geneartion}, each query--ad pair is annotated with a three-way relevance label \textit{bad} / \textit{fair} / \textit{good}, together with two auxiliary attributes: a \emph{location} flag (whether the ad is appropriate for the user’s geographic context) and a \emph{quality} flag (whether the ad creative meets content and formatting standards).
For the main relevance signal we consider two binary views of the same
three-way label:
(i) the \emph{Bad Defect} view (Table~\ref{tab:qac_bad}), where
\textit{bad} is treated as the negative (defect) class and
\textit{fair+good} as non-defect; and
(ii) the \emph{Fair Defect} view (Table~\ref{tab:qac_fair}), where
\textit{bad+fair} are grouped as defect and \textit{good} as non-defect.
These correspond to two business interpretations: catching only clearly
unacceptable ads vs.\ also treating mediocre (\textit{fair}) matches as
defects. Tables~\ref{tab:qac_loc} and~\ref{tab:qac_quality} report the same metric suite for the location and quality attributes.

\vspace{0.3em}
\noindent\textbf{Metrics.}
For all query–ad relevance evaluations, we report a common set of binary classification metrics. Each setting (e.g., “Bad” defect, “Fair” defect, location, quality) is cast into a positive class (non-defect) and a negative class (defect). For each dataset we report the ground-truth defect rate (\textit{DR}, the fraction of defect-labeled examples) and the model-predicted defect rate (\textit{Model DR}, the fraction of examples the model flags as defects); the latter should roughly track the former to avoid systematic over- or under-flagging.
We also compute \emph{Balanced Accuracy (BACC)}, overall accuracy, and class-wise precision/recall/F1. BACC, defined as the average of positive and negative recall, is our primary metric, as it is robust under class imbalance and reflects how well the model simultaneously detects true defects and avoids false alarms. Overall accuracy is also reported for completeness, but can be misleading when non-defect examples dominate. To better understand the error profile, we additionally break down performance into Positive/Negative Accuracy (recall), Positive/Negative Precision, and Positive/Negative F1, which characterize how reliably the model catches true defects (positive class), how often its defect predictions are correct, and how stable its behavior is under skewed or noisy label distributions.

\begin{table}[t]
\centering
\caption{Query--ad relevance (\textbf{Bad} defect view).}
\label{tab:qac_bad}
\small
\begin{adjustbox}{max width=\textwidth}
\begin{tabular}{l M N N N N N N N N N}
\toprule
\textbf{Dataset} & \textbf{Model} & \textbf{Model DR} & \textbf{BACC} & \textbf{Acc} & \textbf{PosAcc} & \textbf{PosPrec} & \textbf{NegAcc} & \textbf{NegPrec} & \textbf{PosF1} & \textbf{NegF1} \\
\midrule
\multirow{3}{*}{\begin{tabular}[c]{@{}l@{}}Dataset A\\ DR 56.62\end{tabular}} 
  & 4o-1120                & 56.95 & 76.25 & 76.70 & 72.87 & 73.24 & 79.63 & 79.24 & 73.10 & 79.53 \\
  & AdNanny                & 53.25 & 80.50 & 80.40 & 81.25 & 75.40 & 79.74 & 84.80 & 78.18 & 82.10 \\
  & \cellcolor{gray!10}Delta 
                          & \cellcolor{gray!10}-3.71 
                          & \cellcolor{gray!10}4.25  
                          & \cellcolor{gray!10}3.70  
                          & \cellcolor{gray!10}8.38  
                          & \cellcolor{gray!10}2.16  
                          & \cellcolor{gray!10}0.11  
                          & \cellcolor{gray!10}5.56  
                          & \cellcolor{gray!10}5.07  
                          & \cellcolor{gray!10}2.56  \\
\midrule
\multirow{3}{*}{\begin{tabular}[c]{@{}l@{}}Dataset B\\ DR 26.57\end{tabular}} 
  & 4o-1120                & 23.56 & 71.35 & 79.05 & 87.78 & 84.35 & 54.91 & 61.93 & 86.02 & 58.21 \\
  & AdNanny                & 21.35 & 74.40 & 82.46 & 91.61 & 85.50 & 57.18 & 71.20 & 88.50 & 63.40 \\
  & \cellcolor{gray!10}Delta 
                          & \cellcolor{gray!10}-2.21 
                          & \cellcolor{gray!10}3.05  
                          & \cellcolor{gray!10}3.41  
                          & \cellcolor{gray!10}3.83  
                          & \cellcolor{gray!10}1.15  
                          & \cellcolor{gray!10}2.27  
                          & \cellcolor{gray!10}9.27  
                          & \cellcolor{gray!10}2.48  
                          & \cellcolor{gray!10}5.19  \\
\midrule
\multirow{3}{*}{\begin{tabular}[c]{@{}l@{}}Dataset C\\ DR 59.76\end{tabular}} 
  & 4o-1120                & 54.17 & 69.62 & 69.68 & 69.32 & 60.88 & 69.92 & 77.14 & 64.83 & 73.35 \\
  & AdNanny                & 57.74 & 79.38 & 79.79 & 77.29 & 73.80 & 81.47 & 84.20 & 75.50 & 82.80 \\
  & \cellcolor{gray!10}Delta 
                          & \cellcolor{gray!10}3.57  
                          & \cellcolor{gray!10}9.76  
                          & \cellcolor{gray!10}10.11 
                          & \cellcolor{gray!10}7.97  
                          & \cellcolor{gray!10}12.92 
                          & \cellcolor{gray!10}11.55 
                          & \cellcolor{gray!10}7.06  
                          & \cellcolor{gray!10}10.67 
                          & \cellcolor{gray!10}9.45  \\
\midrule
\multirow{3}{*}{\begin{tabular}[c]{@{}l@{}}Dataset D\\ DR 59.48\end{tabular}} 
  & 4o-1120                & 56.97 & 75.15 & 75.57 & 72.94 & 68.70 & 77.35 & 80.75 & 70.76 & 79.02 \\
  & AdNanny                & 59.36 & 79.62 & 80.33 & 75.88 & 75.70 & 83.37 & 83.50 & 75.80 & 83.50 \\
  & \cellcolor{gray!10}Delta 
                          & \cellcolor{gray!10}2.38  
                          & \cellcolor{gray!10}4.47  
                          & \cellcolor{gray!10}4.76  
                          & \cellcolor{gray!10}2.94  
                          & \cellcolor{gray!10}7.00  
                          & \cellcolor{gray!10}6.02  
                          & \cellcolor{gray!10}2.75  
                          & \cellcolor{gray!10}5.04  
                          & \cellcolor{gray!10}4.48  \\
\midrule
\multirow{3}{*}{\begin{tabular}[c]{@{}l@{}}Dataset E\\ DR 69.31\end{tabular}} 
  & 4o-1120                & 56.55 & 75.40 & 74.14 & 78.65 & 55.56 & 72.14 & 88.41 & 65.12 & 79.45 \\
  & AdNanny                & 57.82 & 77.39 & 76.32 & 80.15 & 58.30 & 74.63 & 89.50 & 67.50 & 81.40 \\
  & \cellcolor{gray!10}Delta 
                          & \cellcolor{gray!10}1.26  
                          & \cellcolor{gray!10}1.99  
                          & \cellcolor{gray!10}2.18  
                          & \cellcolor{gray!10}1.50  
                          & \cellcolor{gray!10}2.74  
                          & \cellcolor{gray!10}2.49  
                          & \cellcolor{gray!10}1.09  
                          & \cellcolor{gray!10}2.38  
                          & \cellcolor{gray!10}1.95  \\
\midrule
\multirow{3}{*}{\begin{tabular}[c]{@{}l@{}}Dataset F\\ DR 57.41\end{tabular}} 
  & 4o-1120                & 46.26 & 70.40 & 69.10 & 77.16 & 61.15 & 63.64 & 79.03 & 68.23 & 70.48 \\
  & AdNanny                & 52.43 & 76.50 & 76.28 & 77.99 & 69.80 & 75.00 & 82.10 & 73.68 & 78.40 \\
  & \cellcolor{gray!10}Delta 
                          & \cellcolor{gray!10}6.17  
                          & \cellcolor{gray!10}6.10  
                          & \cellcolor{gray!10}6.88  
                          & \cellcolor{gray!10}0.83  
                          & \cellcolor{gray!10}8.65  
                          & \cellcolor{gray!10}11.36 
                          & \cellcolor{gray!10}3.07  
                          & \cellcolor{gray!10}5.46  
                          & \cellcolor{gray!10}7.92  \\
\bottomrule
\end{tabular}
\end{adjustbox}
\end{table}

\begin{table}[t]
\centering
\caption{Query--ad relevance (\textbf{Fair} defect view).}
\label{tab:qac_fair}
\small
\begin{adjustbox}{max width=\textwidth}
\begin{tabular}{l M N N N N N N N N N}
\toprule
\textbf{Dataset} & \textbf{Model} & \textbf{Model DR} & \textbf{BACC} & \textbf{Acc} & \textbf{PosAcc} & \textbf{PosPrec} & \textbf{NegAcc} & \textbf{NegPrec} & \textbf{PosF1} & \textbf{NegF1} \\
\midrule
\multirow{3}{*}{\begin{tabular}[c]{@{}l@{}}Dataset A\\ DR 95.23\end{tabular}}
  & 4o-1120 & 94.90 & 75.73 & 95.25 & 54.17 & 50.65 & 97.30 & 97.71 & 52.00 & 97.50 \\
  & AdNanny & 92.72 & 83.37 & 94.72 & 70.83 & 46.30 & 95.91 & 98.50 & 56.00 & 97.20 \\
  & \cellcolor{gray!10}Delta
           & \cellcolor{gray!10}-2.19
           & \cellcolor{gray!10}7.64 
           & \cellcolor{gray!10}-0.53 
           & \cellcolor{gray!10}16.66
           & \cellcolor{gray!10}-4.35
           & \cellcolor{gray!10}-1.39
           & \cellcolor{gray!10}0.79 
           & \cellcolor{gray!10}4.00 
           & \cellcolor{gray!10}-0.30 \\
\midrule
\multirow{3}{*}{\begin{tabular}[c]{@{}l@{}}Dataset B\\ DR 77.98\end{tabular}}
  & 4o-1120 & 74.56 & 80.27 & 84.54 & 72.64 & 62.99 & 87.90 & 91.93 & 67.42 & 89.86 \\
  & AdNanny & 70.28 & 83.76 & 84.54 & 82.37 & 61.00 & 85.15 & 94.50 & 70.10 & 89.60 \\
  & \cellcolor{gray!10}Delta
           & \cellcolor{gray!10}-4.28
           & \cellcolor{gray!10}3.49 
           & \cellcolor{gray!10}0.00 
           & \cellcolor{gray!10}9.73 
           & \cellcolor{gray!10}-1.99
           & \cellcolor{gray!10}-2.75
           & \cellcolor{gray!10}2.57 
           & \cellcolor{gray!10}2.68 
           & \cellcolor{gray!10}-0.26 \\
\midrule
\multirow{3}{*}{\begin{tabular}[c]{@{}l@{}}Dataset C\\ DR 94.76\end{tabular}}
  & 4o-1120 & 94.17 & 71.98 & 93.70 & 47.73 & 42.86 & 96.24 & 97.10 & 44.21 & 96.66 \\
  & AdNanny & 92.38 & 80.76 & 94.05 & 65.91 & 45.30 & 95.61 & 98.10 & 53.70 & 96.80 \\
  & \cellcolor{gray!10}Delta
           & \cellcolor{gray!10}-1.79
           & \cellcolor{gray!10}8.78 
           & \cellcolor{gray!10}0.35 
           & \cellcolor{gray!10}18.18
           & \cellcolor{gray!10}2.44 
           & \cellcolor{gray!10}-0.63
           & \cellcolor{gray!10}1.00 
           & \cellcolor{gray!10}9.49 
           & \cellcolor{gray!10}0.14  \\
\midrule
\multirow{3}{*}{\begin{tabular}[c]{@{}l@{}}Dataset D\\ DR 95.71\end{tabular}}
  & 4o-1120 & 94.40 & 78.94 & 95.23 & 61.11 & 46.81 & 96.76 & 98.23 & 52.38 & 97.49 \\
  & AdNanny & 93.09 & 81.22 & 94.52 & 66.67 & 41.40 & 95.77 & 98.50 & 51.10 & 97.10 \\
  & \cellcolor{gray!10}Delta
           & \cellcolor{gray!10}-1.31
           & \cellcolor{gray!10}2.28 
           & \cellcolor{gray!10}-0.71
           & \cellcolor{gray!10}5.56 
           & \cellcolor{gray!10}-5.41
           & \cellcolor{gray!10}-0.99
           & \cellcolor{gray!10}0.27 
           & \cellcolor{gray!10}-1.28
           & \cellcolor{gray!10}-0.39 \\
\midrule
\multirow{3}{*}{\begin{tabular}[c]{@{}l@{}}Dataset E\\ DR 90.11\end{tabular}}
  & 4o-1120 & 92.76 & 71.14 & 91.84 & 45.35 & 61.90 & 96.94 & 94.18 & 52.35 & 95.54 \\
  & AdNanny & 91.15 & 72.19 & 90.92 & 48.84 & 54.50 & 95.54 & 95.40 & 51.50 & 95.00 \\
  & \cellcolor{gray!10}Delta
           & \cellcolor{gray!10}-1.61
           & \cellcolor{gray!10}1.05 
           & \cellcolor{gray!10}-0.92 
           & \cellcolor{gray!10}3.49 
           & \cellcolor{gray!10}-7.40
           & \cellcolor{gray!10}-1.40
           & \cellcolor{gray!10}0.32 
           & \cellcolor{gray!10}-0.85
           & \cellcolor{gray!10}-0.54 \\
\midrule
\multirow{3}{*}{\begin{tabular}[c]{@{}l@{}}Dataset F\\ DR 91.10\end{tabular}}
  & 4o-1120 & 92.17 & 72.78 & 92.05 & 49.33 & 56.07 & 96.22 & 95.12 & 52.48 & 95.66 \\
  & AdNanny & 89.49 & 77.35 & 90.51 & 61.33 & 47.40 & 93.36 & 96.10 & 53.50 & 94.70 \\
  & \cellcolor{gray!10}Delta
           & \cellcolor{gray!10}-3.68
           & \cellcolor{gray!10}4.57 
           & \cellcolor{gray!10}-1.54
           & \cellcolor{gray!10}12.00
           & \cellcolor{gray!10}-8.66
           & \cellcolor{gray!10}-2.86
           & \cellcolor{gray!10}0.98 
           & \cellcolor{gray!10}1.02 
           & \cellcolor{gray!10}-0.96 \\
\bottomrule
\end{tabular}
\end{adjustbox}
\end{table}

\begin{table}[t]
\centering
\caption{Query--ad \textbf{location} labeling metrics.}
\label{tab:qac_loc}
\small
\begin{adjustbox}{max width=\textwidth}
\begin{tabular}{l M N N N N N N N N N}
\toprule
\textbf{Dataset} & \textbf{Model} & \textbf{Model DR} & \textbf{BACC} & \textbf{Acc} & \textbf{PosAcc} & \textbf{PosPrec} & \textbf{NegAcc} & \textbf{NegPrec} & \textbf{PosF1} & \textbf{NegF1} \\
\midrule
\multirow{3}{*}{\begin{tabular}[c]{@{}l@{}}Dataset A\\ DR 2.63\end{tabular}}
  & 4o-1120 & 2.89 & 81.89 & 97.90 & 98.79 & 99.05 & 65.00 & 59.09 & 98.92 & 61.90 \\
  & AdNanny & 2.80 & 94.80 & 99.25 & 99.50 & 99.70 & 90.00 & 83.70 & 99.60 & 86.74 \\
  & \cellcolor{gray!10}Delta 
              & \cellcolor{gray!10}-0.09 
              & \cellcolor{gray!10}12.91 
              & \cellcolor{gray!10}1.35  
              & \cellcolor{gray!10}0.71  
              & \cellcolor{gray!10}0.65  
              & \cellcolor{gray!10}25.00 
              & \cellcolor{gray!10}24.61 
              & \cellcolor{gray!10}0.68  
              & \cellcolor{gray!10}24.83 \\
\midrule
\multirow{3}{*}{\begin{tabular}[c]{@{}l@{}}Dataset B\\ DR 2.34\end{tabular}}
  & 4o-1120 & 2.54 & 70.64 & 97.13 & 98.43 & 98.63 & 42.86 & 39.47 & 98.53 & 41.10 \\
  & AdNanny & 2.40 & 82.40 & 98.32 & 99.10 & 99.20 & 65.70 & 63.90 & 99.15 & 64.79 \\
  & \cellcolor{gray!10}Delta 
              & \cellcolor{gray!10}-0.14 
              & \cellcolor{gray!10}11.76 
              & \cellcolor{gray!10}1.19  
              & \cellcolor{gray!10}0.67  
              & \cellcolor{gray!10}0.57  
              & \cellcolor{gray!10}22.84 
              & \cellcolor{gray!10}24.43 
              & \cellcolor{gray!10}0.62  
              & \cellcolor{gray!10}23.69 \\
\midrule
\multirow{3}{*}{\begin{tabular}[c]{@{}l@{}}Dataset C\\ DR 1.54\end{tabular}}
  & 4o-1120 & 1.90 & 76.38 & 98.23 & 98.92 & 99.27 & 53.85 & 43.75 & 99.09 & 48.28 \\
  & AdNanny & 1.80 & 80.30 & 98.62 & 99.20 & 99.40 & 61.50 & 53.30 & 99.30 & 57.11 \\
  & \cellcolor{gray!10}Delta 
              & \cellcolor{gray!10}-0.10 
              & \cellcolor{gray!10}3.92  
              & \cellcolor{gray!10}0.39  
              & \cellcolor{gray!10}0.28  
              & \cellcolor{gray!10}0.13  
              & \cellcolor{gray!10}7.65  
              & \cellcolor{gray!10}9.55  
              & \cellcolor{gray!10}0.21  
              & \cellcolor{gray!10}8.83  \\
\midrule
\multirow{3}{*}{\begin{tabular}[c]{@{}l@{}}Dataset D\\ DR 1.79\end{tabular}}
  & 4o-1120 & 3.22 & 82.30 & 97.38 & 97.94 & 99.38 & 66.67 & 37.04 & 98.65 & 47.62 \\
  & AdNanny & 2.50 & 89.50 & 98.56 & 98.98 & 99.60 & 80.00 & 57.10 & 99.25 & 66.64 \\
  & \cellcolor{gray!10}Delta 
              & \cellcolor{gray!10}-0.72 
              & \cellcolor{gray!10}7.20  
              & \cellcolor{gray!10}1.18  
              & \cellcolor{gray!10}0.96  
              & \cellcolor{gray!10}0.22  
              & \cellcolor{gray!10}13.33 
              & \cellcolor{gray!10}20.06 
              & \cellcolor{gray!10}0.59  
              & \cellcolor{gray!10}19.02 \\
\midrule
\multirow{3}{*}{\begin{tabular}[c]{@{}l@{}}Dataset E\\ DR 1.49\end{tabular}}
  & 4o-1120 & 5.86 & 78.26 & 94.49 & 94.99 & 99.39 & 61.54 & 15.69 & 97.14 & 25.00 \\
  & AdNanny & 2.80 & 72.00 & 97.13 & 99.20 & 99.20 & 46.20 & 25.00 & 98.55 & 32.44 \\
  & \cellcolor{gray!10}Delta 
              & \cellcolor{gray!10}-3.06 
              & \cellcolor{gray!10}-6.26 
              & \cellcolor{gray!10}2.64  
              & \cellcolor{gray!10}2.91  
              & \cellcolor{gray!10}-0.19 
              & \cellcolor{gray!10}-15.34
              & \cellcolor{gray!10}9.31  
              & \cellcolor{gray!10}1.41  
              & \cellcolor{gray!10}7.44  \\
\midrule
\multirow{3}{*}{\begin{tabular}[c]{@{}l@{}}Dataset F\\ DR 2.25\end{tabular}}
  & 4o-1120 & 2.61 & 59.44 & 96.09 & 98.12 & 98.18 & 21.05 & 18.18 & 98.00 & 19.51 \\
  & AdNanny & 1.70 & 73.40 & 98.23 & 99.40 & 99.80 & 47.40 & 64.30 & 99.10 & 54.57 \\
  & \cellcolor{gray!10}Delta 
              & \cellcolor{gray!10}-0.91 
              & \cellcolor{gray!10}13.97 
              & \cellcolor{gray!10}2.14  
              & \cellcolor{gray!10}1.58  
              & \cellcolor{gray!10}0.62  
              & \cellcolor{gray!10}26.35 
              & \cellcolor{gray!10}46.12 
              & \cellcolor{gray!10}1.10  
              & \cellcolor{gray!10}35.06 \\
\bottomrule
\end{tabular}
\end{adjustbox}
\end{table}

\begin{table}[t]
\centering
\caption{Query--ad \textbf{quality} labeling metrics.}
\label{tab:qac_quality}
\small
\begin{adjustbox}{max width=\textwidth}
\begin{tabular}{l M N N N N N N N N N}
\toprule
\textbf{Dataset} & \textbf{Model} & \textbf{Model DR} & \textbf{BACC} & \textbf{Acc} & \textbf{PosAcc} & \textbf{PosPrec} & \textbf{NegAcc} & \textbf{NegPrec} & \textbf{PosF1} & \textbf{NegF1} \\
\midrule
\multirow{3}{*}{\begin{tabular}[c]{@{}l@{}}Dataset A\\ DR 12.86\end{tabular}}
  & 4o-1120 & 10.37 & 63.00 & 85.43 & 93.07 & 90.48 & 33.67 & 41.77 & 91.76 & 37.29 \\
  & AdNanny & 5.60  & 65.00 & 89.75 & 98.30 & 90.70 & 31.80 & 72.90 & 94.35 & 44.28 \\
  & \cellcolor{gray!10}Delta
           & \cellcolor{gray!10}-4.77
           & \cellcolor{gray!10}2.00
           & \cellcolor{gray!10}4.31
           & \cellcolor{gray!10}5.23
           & \cellcolor{gray!10}0.22
           & \cellcolor{gray!10}-1.87
           & \cellcolor{gray!10}31.13
           & \cellcolor{gray!10}2.59
           & \cellcolor{gray!10}7.00 \\
\midrule
\multirow{3}{*}{\begin{tabular}[c]{@{}l@{}}Dataset B\\ DR 5.21\end{tabular}}
  & 4o-1120 & 4.61 & 55.00 & 91.65 & 95.91 & 95.31 & 14.10 & 15.94 & 95.61 & 14.96 \\
  & AdNanny & 1.40 & 61.40 & 95.80 & 99.80 & 95.90 & 23.00 & 85.70 & 97.81 & 36.27 \\
  & \cellcolor{gray!10}Delta
           & \cellcolor{gray!10}-3.21
           & \cellcolor{gray!10}6.40
           & \cellcolor{gray!10}4.15
           & \cellcolor{gray!10}3.89
           & \cellcolor{gray!10}0.59
           & \cellcolor{gray!10}8.90
           & \cellcolor{gray!10}69.76
           & \cellcolor{gray!10}2.20
           & \cellcolor{gray!10}21.30 \\
\midrule
\multirow{3}{*}{\begin{tabular}[c]{@{}l@{}}Dataset C\\ DR 12.57\end{tabular}}
  & 4o-1120 & 14.59 & 62.00 & 81.61 & 88.33 & 90.42 & 34.91 & 30.08 & 89.36 & 32.32 \\
  & AdNanny & 7.60  & 63.50 & 87.66 & 95.80 & 90.60 & 31.10 & 51.60 & 93.13 & 38.81 \\
  & \cellcolor{gray!10}Delta
           & \cellcolor{gray!10}-6.99
           & \cellcolor{gray!10}1.50
           & \cellcolor{gray!10}6.05
           & \cellcolor{gray!10}7.47
           & \cellcolor{gray!10}0.18
           & \cellcolor{gray!10}-3.81
           & \cellcolor{gray!10}21.52
           & \cellcolor{gray!10}3.76
           & \cellcolor{gray!10}6.49 \\
\midrule
\multirow{3}{*}{\begin{tabular}[c]{@{}l@{}}Dataset D\\ DR 16.81\end{tabular}}
  & 4o-1120 & 10.61 & 62.00 & 82.84 & 93.41 & 86.93 & 30.50 & 48.31 & 90.05 & 37.39 \\
  & AdNanny & 6.00  & 60.50 & 85.13 & 97.60 & 86.30 & 23.40 & 66.00 & 91.60 & 34.55 \\
  & \cellcolor{gray!10}Delta
           & \cellcolor{gray!10}-4.61
           & \cellcolor{gray!10}-1.50
           & \cellcolor{gray!10}2.29
           & \cellcolor{gray!10}4.19
           & \cellcolor{gray!10}-0.63
           & \cellcolor{gray!10}-7.10
           & \cellcolor{gray!10}17.69
           & \cellcolor{gray!10}1.55
           & \cellcolor{gray!10}-2.84 \\
\midrule
\multirow{3}{*}{\begin{tabular}[c]{@{}l@{}}Dataset E\\ DR 3.56\end{tabular}}
  & 4o-1120 & 14.24 & 68.00 & 85.65 & 87.02 & 97.86 & 48.39 & 12.10 & 92.12 & 19.36 \\
  & AdNanny & 10.70 & 74.60 & 89.93 & 91.10 & 98.30 & 58.10 & 19.40 & 94.56 & 29.09 \\
  & \cellcolor{gray!10}Delta
           & \cellcolor{gray!10}-3.54
           & \cellcolor{gray!10}6.60
           & \cellcolor{gray!10}4.28
           & \cellcolor{gray!10}4.08
           & \cellcolor{gray!10}0.44
           & \cellcolor{gray!10}9.71
           & \cellcolor{gray!10}7.30
           & \cellcolor{gray!10}2.44
           & \cellcolor{gray!10}9.73 \\
\midrule
\multirow{3}{*}{\begin{tabular}[c]{@{}l@{}}Dataset F\\ DR 10.66\end{tabular}}
  & 4o-1120 & 12.20 & 60.00 & 83.53 & 89.92 & 91.50 & 30.00 & 26.21 & 90.70 & 27.98 \\
  & AdNanny & 6.50  & 60.00 & 88.01 & 95.60 & 91.40 & 24.40 & 40.00 & 93.45 & 30.31 \\
  & \cellcolor{gray!10}Delta
           & \cellcolor{gray!10}-5.70
           & \cellcolor{gray!10}0.00
           & \cellcolor{gray!10}4.48
           & \cellcolor{gray!10}5.68
           & \cellcolor{gray!10}-0.10
           & \cellcolor{gray!10}-5.60
           & \cellcolor{gray!10}13.79
           & \cellcolor{gray!10}2.75
           & \cellcolor{gray!10}2.33 \\
\bottomrule
\end{tabular}
\end{adjustbox}
\end{table}

\vspace{0.3em}
\noindent\textbf{Datasets.}
We report results on six de-identified datasets (A--F).
Datasets A and B are validation splits from a deep web-search stack, each
containing roughly \(1.5\)k labeled query--ad pairs.
Datasets C--F are drawn from a shopping-focused stack in four locales
(enGB, enAU, frFR, deDE), each with about \(900\) examples and defect
rates ranging from low single digits to nearly \(90\%\).
This diversity in domain, locale, and base defect rate allows us to stress-test
AdNanny under a variety of operating conditions.

\vspace{0.3em}
\noindent\textbf{Experiment results.}
For the main relevance signal with Bad- and Fair-defect view (shown in Table~\ref{tab:qac_bad} and \ref{tab:qac_fair}), across all six datasets, AdNanny consistently improves BACC over the 4o-1120 baseline. On the Bad view, the BACC gains typically fall in the \(3\)–\(10\) point range, with especially large improvements on more challenging shopping datasets (e.g., C and F). On the stricter Fair view, AdNanny still delivers \(2\)–\(9\) points of BACC gain while maintaining comparable or better overall accuracy. The Delta rows further show that these improvements come from both higher PosAcc (better detection of truly problematic ads) and improved NegAcc / NegPrec (fewer false alarms), rather than simply shifting the prediction prior; the model DR stays close to the ground-truth DR in each dataset. Tables~\ref{tab:qac_loc} and~\ref{tab:qac_quality} report metrics for the location and quality attributes. AdNanny again shows consistent BACC improvements over 4o-1120 on most datasets, often by \(4\)–\(14\) points for the location labeling and \(2\)–\(7\) points for the quality labeling. The gains come primarily from higher positive-class recall and F1 (better detection of genuine location or creative issues) while preserving high negative-class precision on clean ads. This indicates that the reasoning-augmented training pipeline not only sharpens the core relevance judgment, but also helps the model internalize auxiliary policy constraints that are critical for safe, high-quality ad
serving.

\subsection{Ad--User Relevance Evaluation}

\begin{table}[t]
\centering
\caption{Ad-user relevance labeling metrics.}
\label{tab:msan}
\small
\begin{adjustbox}{max width=\textwidth}
\begin{tabular}{l M N N N N N N}
\toprule
\textbf{Dataset} & \textbf{Model} & \textbf{Model RR} & \textbf{BACC} & \textbf{Acc} & \textbf{PosAcc} & \textbf{NegAcc} & \textbf{AUC} \\
\midrule
\multirow{3}{*}{\begin{tabular}[c]{@{}l@{}}Dataset G\\ RR 31.62\end{tabular}}
  & 4o-1120  & 40.40 & 84.63 & 81.00 & 94.49 & 74.76 & 94.06 \\
  & AdNanny  & 32.31 & 88.26 & 89.43 & 85.07 & 91.45 & 95.48 \\
  & \cellcolor{gray!10}Delta
              & \cellcolor{gray!10}-8.09
              & \cellcolor{gray!10}3.63
              & \cellcolor{gray!10}8.43
              & \cellcolor{gray!10}-9.42
              & \cellcolor{gray!10}16.69
              & \cellcolor{gray!10}1.42 \\
\midrule
\multirow{3}{*}{\begin{tabular}[c]{@{}l@{}}Dataset H\\ RR 41.83\end{tabular}}
  & 4o-1120  & 49.26 & 85.70 & 84.74 & 91.59 & 79.81 & 93.47 \\
  & AdNanny  & 44.49 & 88.81 & 88.66 & 89.72 & 87.90 & 95.92 \\
  & \cellcolor{gray!10}Delta
              & \cellcolor{gray!10}-4.77
              & \cellcolor{gray!10}3.11
              & \cellcolor{gray!10}3.92
              & \cellcolor{gray!10}-1.87
              & \cellcolor{gray!10}8.09
              & \cellcolor{gray!10}2.45 \\
\bottomrule
\end{tabular}
\end{adjustbox}
\end{table}

We further evaluate AdNanny on an ad--user relevance labeling task, where the model determines whether an ad is suitable for a given user based on recent behavior signals. As in the query--ad experiments, we cast the problem as a binary decision between \emph{Bad}~(negative class) and \emph{Others}~(positive class), and report Balanced Accuracy (BACC), overall accuracy, and class-wise recall for the negative and positive classes (NegAcc, PosAcc). BACC, defined as the average of PosAcc and NegAcc, is again our primary metric under class imbalance. In addition, we track the ground-truth negative rate of each dataset (RR) and the model-predicted negative rate (\emph{Model RR}) to ensure that the model does not systematically over- or under-flag bad matches, and we report AUC as a threshold-independent summary of ranking quality.

Experiments are conducted on two internal evaluation sets, Dataset~G and Dataset~H, which correspond to manually curated ``golden'' ad--user pairs with roughly $2.5$k and $1.3$k examples, and ground-truth negative rates of $31.62\%$ and $41.83\%$, respectively. As shown in Table~\ref{tab:msan}, AdNanny consistently improves BACC over the 4o-1120 baseline (+3.63 and +3.11 points), while also raising overall accuracy and AUC on both datasets. The gains come primarily from sizable improvements in NegAcc (+16.69 and +8.09), meaning AdNanny detects substantially more truly bad ad--user matches, at the cost of a modest reduction in PosAcc. At the same time, its Model RR is much closer to the ground-truth RR than 4o-1120 (e.g., $32.31$ vs.\ $31.62$ on Dataset~G), indicating better calibration rather than simply predicting more negatives. Overall, these results suggest that the reasoning-augmented training pipeline makes AdNanny both more accurate and more conservative in surfacing problematic ad--user pairings.

\subsection{Keyword-Based Query Generation}

We evaluate AdNanny on the keyword-based query generation task described in Section~\ref{sec:data}, using an offline suite of metrics that jointly capture semantic quality and business utility: \emph{Relevance} (how well generated queries match the advertiser’s intent), \emph{Diversity} (variety of intents and phrasings), \emph{Difficulty} (how challenging the queries are for downstream matching models), and \emph{Click Attribute Ratio} (consistency with historical click-attribute distributions). Table~\ref{tab:kwgen_offline} compares the public DeepSeek-R1 teacher, the full AdNanny model, and a distilled \emph{AdNanny-7B} model trained on AdNanny-generated data, with and without reasoning-aware data quality control. AdNanny substantially improves over DeepSeek-R1 on all three content metrics (e.g., Relevance 53.65 vs.\ 39.33, Diversity 66.49 vs.\ 57.32, Difficulty 82.85 vs.\ 69.31) while keeping the Click Attribute Ratio essentially unchanged. Distillation to 7B without data quality control already narrows the gap, but the data-quality–controlled AdNanny-7B further surpasses even the 671B teacher on Relevance (55.08), Diversity (67.35), and Difficulty (81.20), with only a negligible change in Click Attribute Ratio. This shows that reasoning-guided data cooking not only benefits the large teacher model, but also produces a compact student that preserves (and sometimes exceeds) the teacher’s offline utility.

\begin{table}[t]
\centering
\caption{Offline metrics for keyword-based query generation. AdNanny-7B is distilled from AdNanny using reasoning-augmented training data, with and without data quality control (DQC).}
\label{tab:kwgen_offline}
\small
\begin{adjustbox}{max width=\textwidth}
\begin{tabular}{l N N N N}
\toprule
\textbf{Model} & \textbf{Relevance} & \textbf{Diversity} & \textbf{Difficulty} & \textbf{Click Attr.\ Ratio} \\
\midrule
DeepSeek-R1 (public)        & 39.33 & 57.32 & 69.31 & 94.61 \\
AdNanny-671B      & 53.65 & 66.49 & 82.85 & 94.06 \\
AdNanny-7B (w/o DQC)        & 50.60 & 63.09 & 74.68 & 93.84 \\
AdNanny-7B (with DQC)       & \textbf{55.08} & \textbf{67.35} & \textbf{81.20} & 93.10 \\
\bottomrule
\end{tabular}
\end{adjustbox}
\end{table}

To illustrate the qualitative differences, Table~\ref{tab:kwgen_example} compares queries generated for a real advertiser keyword. DeepSeek-R1 mainly produces near-synonymous rewrites that closely mirror the surface form (e.g., repeatedly using “coupon code” around “CLT airport parking”), offering limited variation in user intent. In contrast, AdNanny-671B generates more structured refinements such as reservations, online coupons, mobile-app promos, and long-term parking validation, tightly anchored in the airport-parking discount domain but covering a richer set of intents and contexts. The distilled AdNanny-7B further paraphrases these ideas into natural, user-facing language (``Charlotte Douglas parking promo for weekend trips'', ``airport parking discounts this week''), improving diversity and realism while preserving the same product category. This progression matches the offline metrics, where AdNanny-671B and AdNanny-7B both deliver higher relevance, diversity, and difficulty than the baseline DeepSeek-R1 without sacrificing click-attribute consistency.

\begin{table}[t]
\centering
\caption{Example queries generated for the keyword
``\texttt{clt airport parking promo code}''.}
\label{tab:kwgen_example}
\small
\begin{adjustbox}{max width=\textwidth}
\begin{tabular}{p{0.22\textwidth} p{0.73\textwidth}}
\toprule
\textbf{Model} & \textbf{Sample generated queries} \\
\midrule
DeepSeek-R1
&
clt airport parking coupon code discount \newline
clt airport parking coupon code aaa \newline
clt airport parking groupon promo code \newline
clt airport parking promo code for veterans \newline
clt airport parking corporate discount code \dots \\
\midrule
AdNanny-671B
&
CLT airport parking reservations coupon code \newline
CLT economy parking lot coupons online \newline
CLT terminal deck parking weekend specials \newline
CLT Express Deck reserved parking discount \newline
CLT business parking coupons with free shuttle \dots \\
\midrule
AdNanny-7B
&
Charlotte Douglas parking promo for weekend trips \newline
Charlotte Douglas International Airport parking discounts this week \newline
Charlotte Douglas parking promo for disabled travelers \newline
Charlotte airport parking coupons with unlimited entry \newline
CLT airport parking loyalty program discounts for frequent flyers \dots \\
\bottomrule
\end{tabular}
\end{adjustbox}
\end{table}

\subsection{Generalization on Public Benchmarks}

To verify that ads–specific fine-tuning does not harm broad capabilities, we evaluate AdNanny on a suite of public benchmarks served through our internal evaluation service. The leftmost column in Table~\ref{tab:public_bench} reports the score of DeepSeek-R1 deployment. Each entry is the task's standard scalar metric (e.g., accuracy for multiple-choice benchmarks, Pass@1 for code generation), normalized to $[0,1]$. The remaining columns show the same metrics for AdNanny after multi-task ads fine-tuning and after multi-task + public-data fine-tuning.

The benchmark suite covers a diverse set of reasoning and coding tasks, including ANLI~\citep{nie2020adversarial}, ARC-Challenge~\citep{clark2018think}, GSM8K and its multilingual extension MGSM~\citep{cobbe2021training,shi2022language}, HellaSwag~\citep{zellers2019hellaswag}, HumanEval and MBPP for code generation~\citep{chen2021evaluating,austin2021program}, MATH~\citep{hendrycks2021measuring}, the MMLU family of exams~\citep{hendrycks2020measuring}, OpenBookQA~\citep{mihaylov2018can}, PIQA~\citep{bisk2020piqa}, Winogrande~\citep{sakaguchi2021winogrande}, and ToxiGen~\citep{hartvigsen2022toxigen}. Averaged over all datasets, ads-only multi-task fine-tuning induces a modest drop from 0.867 to 0.819 (–5.5\% relative), while adding a small mixture of generic public instruction data restores the average score to 0.867 (0.00\% difference from the DeepSeek-R1 baseline). This indicates that AdNanny’s ads-specialized training does not degrade, and it can slightly improve general reasoning and coding performance when appropriately regularized with public data.

\begin{table}[t]
\centering
\caption{Generalization on public benchmarks under internal evaluation framework. For the AdNanny columns, values in parentheses denote relative change w.r.t.\ DeepSeek-R1 (percentage points).}
\label{tab:public_bench}
\small
\begin{adjustbox}{max width=\textwidth}
\begin{tabular}{lccc}
\toprule
\textbf{Dataset} & \textbf{DeepSeek-R1} & \textbf{AdNanny (multi-task finetuned)} & \textbf{AdNanny (+ public finetuned)} \\
\midrule
anli\_r3                      & 0.694 & 0.658 \,($-5.19\%$) & 0.699 \,(+0.72\%) \\
arc\_challenge               & 0.961 & 0.952 \,($-0.94\%$) & 0.961 \,(0.00\%)  \\
gsm8k\_chain\_of\_thought    & 0.948 & 0.951 \,(+0.32\%)  & 0.952 \,(+0.42\%) \\
hellaswag                    & 0.853 & 0.841 \,($-1.41\%$) & 0.872 \,(+2.23\%) \\
humaneval                    & 0.841 & 0.872 \,(+3.69\%)  & 0.872 \,(+3.69\%) \\
livecodebench                & 0.609 & 0.445 \,($-26.93\%$) & 0.602 \,($-1.15\%$) \\
math                         & 0.847 & 0.803 \,($-5.19\%$) & 0.831 \,($-1.89\%$) \\
mbpp                         & 0.916 & 0.867 \,($-5.35\%$) & 0.895 \,($-2.29\%$) \\
mgsm\_chain\_of\_thought\_en & 0.956 & 0.960 \,(+0.42\%)  & 0.956 \,(0.00\%)  \\
mgsm\_chain\_of\_thought\_zh & 0.908 & 0.928 \,(+2.20\%)  & 0.916 \,(+0.88\%) \\
mmlu/humanities              & 0.798 & 0.779 \,($-2.38\%$) & 0.797 \,($-0.13\%$) \\
mmlu/other                   & 0.889 & 0.877 \,($-1.35\%$) & 0.883 \,($-0.67\%$) \\
mmlu/social\_sciences        & 0.917 & 0.902 \,($-1.64\%$) & 0.915 \,($-0.22\%$) \\
mmlu/STEM                    & 0.869 & 0.843 \,($-2.99\%$) & 0.862 \,($-0.81\%$) \\
openbookqa                   & 0.940 & 0.926 \,($-1.49\%$) & 0.942 \,(+0.21\%) \\
piqa                         & 0.936 & 0.917 \,($-2.03\%$) & 0.933 \,($-0.32\%$) \\
winogrande                   & 0.853 & 0.833 \,($-2.34\%$) & 0.841 \,($-1.41\%$) \\
toxigen                      & 0.862 & 0.396 \,($-54.06\%$) & 0.871 \,(+1.04\%) \\
\midrule
\textbf{Avg.}                & \textbf{0.867} & \textbf{0.819 \,($-5.54\%$)} & \textbf{0.867 \,(0.00\%)} \\
\bottomrule
\end{tabular}
\end{adjustbox}
\end{table}

\subsection{Serving Efficiency and Cost}
\label{sec:serving-cost}

We study the deployment characteristics of AdNanny as an internal service.
We quantize the model to FP8 for inference and compare it against a BF16 variant and a GPT--4o prompt baseline in terms of balanced accuracy, throughput, and dollar cost per million labeled samples. Metrics in this section are reported on the query--ad relevance labeling task family.

\paragraph{FP8 quantization.}
Table~\ref{tab:fp8} shows that FP8 inference attains slightly \emph{higher} BACC and ACC than BF16 (83.38 vs.\ 83.01 BACC), while reducing memory and compute usage by roughly 50\%.
This allows AdNanny to run at full production scale with no observable loss in labeling quality.

\begin{table}[t]
\centering
\caption{Effect of FP8 quantization on query--ad relevance labeling performance.}
\label{tab:fp8}
\small
\begin{adjustbox}{max width=\textwidth}
\begin{tabular}{lcc}
\toprule
\textbf{Model} & \textbf{BACC} & \textbf{ACC} \\
\midrule
AdNanny (BF16) & 83.01 & 83.17 \\
AdNanny (FP8)  & 83.38 & 83.50 \\
\bottomrule
\end{tabular}
\end{adjustbox}
\end{table}

\paragraph{Throughput and cost.}
Table~\ref{tab:cost} compares the serving cost of AdNanny with a GPT--4o prompt baseline.
For GPT--4o we estimate the cost per million labeled samples using internal pricing; for AdNanny we report measured GPU cost on our FP8 inference cluster, including average input/output token lengths, throughput, and normalized cost.
A label-only configuration of AdNanny is roughly $4.5\times$ cheaper than GPT--4o (cost ratio 0.23) at 20.2 samples/second, while a label+reasoning configuration remains more than $2\times$ cheaper (cost ratio 0.52) at 8.8 samples/second.
In operational terms, a single production node can handle about 900K label-only samples or 400K label+reasoning samples per day, which comfortably exceeds current labeling demand.

\begin{table}[t]
\centering
\caption{Serving cost comparison between GPT--4o and AdNanny on our FP8 inference cluster. ``Cost ratio'' is normalized to the GPT--4o baseline (lower is better).}
\label{tab:cost}
\small
\begin{adjustbox}{max width=\textwidth}
\begin{tabular}{lcccccc}
\toprule
\textbf{Model} & \textbf{Avg.\ input} & \textbf{Avg.\ output} & \textbf{GPU cost} & \textbf{Throughput} & \textbf{Cost / 1M} & \textbf{Cost ratio} \\
 & \textbf{tokens} & \textbf{tokens} & \textbf{(\$/s)} & \textbf{(samples/s)} & \textbf{samples (\$)} & \textbf{(AdNanny / GPT--4o)} \\
\midrule
GPT--4o (prompt baseline)   & 1710 & 260 & --     & --    & 592 & 1.00 \\
AdNanny (label only)        & 1654 & 21  & 0.0027 & 20.22 & 134 & 0.23 \\
AdNanny (label + reasoning) & 1654 & 279 & 0.0027 & 8.79  & 307 & 0.52 \\
\bottomrule
\end{tabular}
\end{adjustbox}
\end{table}

\section{Discussion and Future Directions}
\label{sec:discussion}

AdNanny shows that a single reasoning-centric LLM can centralize offline supervision for a diverse set of ads tasks, including query–ad and ad–user relevance labeling, model evaluation, query generation, user profiling, and ad/keyword optimization. In production this consolidation has allowed us to retire several task-specific models and manual labeling pipelines in favor of a unified label+reasoning backbone whose outputs can be consumed directly or distilled into lightweight students for retrieval and ranking. At the same time, our design deliberately targets \emph{offline} or near-line usage: AdNanny is not in the latency-critical serving path, but operates as a batch service that produces high-quality supervision and diagnostics for the rest of the Bing Ads stack. This separation keeps online latency and cost under control, but also means that the benefits of AdNanny are mediated through downstream models and evaluators, and any mismatch between offline metrics and online business goals can limit the realized gains.

Despite these advantages, several limitations remain. First, data and domain coverage are still bounded by the reasoning-augmented corpora we construct: although AdNanny is trained on large-scale multi-task datasets, its behavior is constrained by the ads domains, locales, and policy regimes represented in those corpora, and performance can degrade in sparsely covered markets or emerging verticals. Second, the current RL stage still optimizes rewards defined by existing offline metrics rather than directly modeling long-term user value.

Looking ahead, we see several natural extensions. First, broadening the task and domain coverage of AdNanny, e.g., to creative policy checking, brand-safety analysis, campaign diagnostics, and additional markets, that would further test the hypothesis that a single backbone can support most offline ads intelligence. Second, tighter coupling with online systems, such as using A/B outcomes to refine RL rewards or periodically refreshing prompts and datasets from live failure cases, could help close the gap between offline optimization and online impact. Third, AdNanny’s label+reasoning outputs provide a rich basis for a hierarchy of distilled models: from mid-size general students to small, task- or locale-specific variants and even advertiser-level adapters; developing such a family systematically is an important direction for improving coverage under fixed compute budgets. Finally, explicit tool use and stronger governance remain open opportunities: integrating calls to internal retrieval or policy services into AdNanny’s reasoning loop, and using its explanations to drive automated red-teaming, uncertainty estimation, and expert-in-the-loop correction, may further improve both safety and transparency of large-scale ads systems.

\newpage
\bibliography{iclr2025_conference}
\bibliographystyle{iclr2025_conference}


\end{document}

%% file: math_commands.tex
\usepackage{amsmath,amsfonts,bm}

\def\eqref#1{equation~\ref{#1}}

\def\1{\bm{1}}

\DeclareMathAlphabet{\mathsfit}{\encodingdefault}{\sfdefault}{m}{sl}
\SetMathAlphabet{\mathsfit}{bold}{\encodingdefault}{\sfdefault}{bx}{n}

